\documentclass[twocolumn, pra, showpacs]{revtex4}
\usepackage{amssymb}
\usepackage{graphicx}
\usepackage{dcolumn}
\usepackage{bm}
\usepackage{amsmath}
\usepackage{textcomp}
\usepackage[dvipdfm,colorlinks,linkcolor=red,citecolor=blue]{hyperref}

\begin{document}

\title{Collective excitation of a trapped Bose-Einstein condensate with
spin-orbit coupling}
\author{Li Chen$^{1,2}$}
\author{Han Pu$^{2,3}$}
\email{hpu@rice.edu}
\author{Zeng-Qiang Yu$^1$}
\author{Yunbo Zhang$^1$}
\email{ybzhang@sxu.edu.cn}
\affiliation{$^1${Institute of Theoretical Physics, Shanxi University, Taiyuan, Shanxi
030006, P. R. China}\\
$^2${Department of Physics and Astronomy, and Rice Center for Quantum
Materials, Rice University, Houston, TX 77005, USA}\\
$^3${Center for Cold Atom Physics, Chinese Academy of Sciences, Wuhan
430071, China}}

\begin{abstract}
We investigate the collective excitations of a Raman-induced spin-orbit
coupled Bose-Einstein condensate confined in a quasi one-dimension harmonic
trap using the Bogoliubov method. By tuning the Raman coupling strength,
three phases of the system can be identified. By calculating the transition
strength, we are able to classify various excitation modes that are
experimentally relevant. We show that the three quantum phases possess
distinct features in their collective excitation properties. In particular,
the spin dipole and the spin breathing modes can be used to clearly map out
the phase boundaries. We confirm these predictions by direct numerical
simulations of the quench dynamics that excites the relevant collective
modes.
\end{abstract}

\maketitle

\section{INTRODUCTION}

In recent years, an important breakthrough in cold atom physics is the
realization of spin-orbit (SO) coupling \cite{Goldman2014, Zhai2015,
Lin2011, Wang2012, Cheuk2012}. The SO coupled Bose gases, which has no analog in conventional solid materials which deal with fermionic systems, present rich many-body quantum phases such as stripe phase \cite%
{Ho2011, YLi2012, YLi2013, SJi2014, JLi2016} and skyrmion lattices \cite%
{Spintexture}. However, despite of the tremendous attention they have
attracted, direct evidences of these exotic phases are still lacking \cite%
{note}.

The main purpose of the present work is to show that different phases of an
SO coupled Bose-Einstein condensate (BEC) features distinctive collective
excitations, which can therefore be used to distinguish various phases. We
will focus on the most commonly achieved system: a spin-1/2 BEC confined in
a harmonic trap and subject to the Raman-induced equal-weight Rashba
Dresselhaus SO coupling. Some of the nontrivial properties of the collective
excitation of such a system has already been explored. For example, the
softening of roton-like gap as well as the sound velocity are experimentally
measured \cite{Martone2012,UniformExcitationT}, and the deviation of the
dipole oscillation frequency away from the trapping frequency has also been
observed \cite{JZhang2012, ZChen2012, YLi2012SumRule, Stringari2016}. This
deviation is a distinct feature of the SO coupling. Here we examine the
Bogoliubov spectrum of the system, develop a technique to classify different
types of collective excitations across the whole phase diagram, and show how
they can help us identify different phases.

\section{MODEL}

We consider an effectively one-dimension system by assuming that the Raman
beams propagate along the $x$-axis with vanishing two-photon detuning, and
the BEC is tightly confined along the $y$- and the $z$-axis. Under the
mean-field framework, the BEC is governed by the following Gross-Pitaeviskii
(GP) equation: (we set $\hbar = M = 1$ with $M$ being the atomic mass)%
\begin{equation}
i{\partial \mathbf{\Psi }\left( x,t\right) }/{\partial t}=\left( H_{0}+%
\mathcal{G}\right) \mathbf{\Psi }\left( x,t\right) \,,  \label{B1}
\end{equation}
where we have labeled the two spin components as $\uparrow$ and $\downarrow$%
, $\mathbf{\Psi }=\left( \psi _{\uparrow },\psi _{\downarrow }\right) ^{T}$
is the spinor wave function which is normalized such that $\int dx|\mathbf{%
\Psi }|^2=N$ with $N$ being the total atom number,
\begin{equation}
H_{0}= k_{x} ^{2}/2-k_{L}k_{x}\sigma _{z}+\Omega \sigma _{x}/2+V\left(
x\right) \,,  \label{h0}
\end{equation}
is the single-particle Hamiltonian where $k_{L}$ denotes the Raman recoil
momentum, $V=\omega _{x}^{2}x^{2}/2$ represents the external harmonic
potential with $\omega _{x}$ being the trapping frequency, and $\Omega $
denotes the Raman coupling strength. In Eq.~(\ref{B1}), $\mathcal{G}%
=diag\left( g_{\uparrow \uparrow }\left\vert \psi _{\uparrow }\right\vert
^{2}+g_{\uparrow \downarrow }\left\vert \psi _{\downarrow }\right\vert
^{2},g_{\downarrow \downarrow }\left\vert \psi _{\downarrow }\right\vert
^{2}+g_{\uparrow \downarrow }\left\vert \psi _{\uparrow }\right\vert
^{2}\right) $ characterizes the two-body interaction. Here we assume that
the interaction is repulsive such that all interaction strengths $g_{\sigma
\sigma^{\prime }}>0$. Moreover, for simplicity, we will take the intra-spin
interaction to be equal, i.e., $g_{\uparrow \uparrow}=g_{\downarrow
\downarrow}=g$.

For a homogeneous system, the single-particle ground state is doubly
degenerate, occurring at $k_x= \pm k_0 = \pm \sqrt{k_L^2-\Omega^2/4k_L^2}$
when $\Omega < 4E_L$, and for $\Omega > 4E_L$, the two degenerate states
merge into a single one with $k_0=0$. Here, $E_{L}\equiv k_{L}^{2}/2$ is the
recoil energy. This leads to three mean-field BEC phases: For $\Omega > 4E_L$%
, all atoms condense to the zero momentum state and hence this phase is
termed as zero-momentum phase (ZM); for $\Omega < 4E_L$, depending on the
interaction strength, we may have all the atoms condense to one of the
degenerate single-particle ground states and we have the plane-wave phase
(PW); or the atoms can condense to an equal-weight superposition of the two
degenerate single-particle ground states and we have the stripe phase (ST),
as both spin components exhibit density stripes. In the presence of a weak
harmonic trap, even though momentum is no longer a good quantum number, the
analog of all the three phases can still be easily identified, and the main
effects of the trap is to provide an overall envelop for the atomic density
and the boundaries between different phases are shifted \cite{zhu}.

We obtain the condensate ground state wave function $\mathbf{\Psi }_{0}$ by
numerically propagating the GP equation in imaginary time \cite{Antoine2014}%
. The numerical results are in very good agreement with the following ans%
\"{a}ts \cite{YLi2012}
\begin{equation}
\mathbf{\Psi }_{0}\approx \left[ C_{1}\left(
\begin{array}{c}
\cos \theta \\
-\sin \theta%
\end{array}%
\right) e^{ik_{0}x}+C_{2}\left(
\begin{array}{c}
\sin \theta \\
-\cos \theta%
\end{array}%
\right) e^{-ik_{0}x}\right] G.  \label{B2}
\end{equation}%
Here, $G\left( x\right) = e^{-x^{2}/2w^{2}}$ is a Gaussian envelop
accounting for the trap confinement with $w$ being the envelop width, $C_{1}$
and $C_{2}$ are amplitudes for the two plane waves with momentum $\pm k_{0}$%
, respectively. The three phases can then be characterized as follows. For
the ST phase, we have $C_1=C_2$ and $k_0\neq 0$; for the PW phase, $C_1=0$
or $C_2=0$ and $k_0 \neq0$; for the ZM phase, $C_1=C_2$ and $k_0=0$.

Different phases can be accessed by tuning the Raman coupling strength $%
\Omega$. For strong inter-spin interaction with $g_{\uparrow \downarrow} \ge
g$, the system is in the PW (ZM) phase if $\Omega <\Omega _{C}^{P-Z}$ ($%
\Omega > \Omega _{C}^{P-Z}$) and the ST phase is absent. The critical value $%
\Omega _{C}^{P-Z}=4E_L$ for a homogeneous system, and is slightly down
shifted by the trap \cite{zhu}. For weaker inter-spin interaction with $%
g_{\uparrow \downarrow} < g$, the ST phase is also present at small Raman
coupling strength $\Omega < \Omega _{C}^{S-P}$, the PW phase exists when $%
\Omega _{C}^{S-P}<\Omega <\Omega _{C}^{P-Z}$, and the ZM phase remains at
large Raman coupling strength when $\Omega >\Omega _{C}^{P-Z}$. Previous
studies have shown that the ST to PW transition at $\Omega _{C}^{S-P}$ is of
first-order, whereas the PW to ZM transition at $\Omega _{C}^{P-Z}$ is of
second-order \cite{YLi2012}.

In our calculation, we will take $g_{\uparrow \downarrow}=0.7g$. As in this
case, all three phases are present. In addition, we will take the atom
number to be $N=2000$, and consider a relatively weak but realistic trap
with $\omega _{x}=0.02E_{L}$. For this set of parameters, the two critical
Raman coupling strengths are $\Omega_{C}^{S-P} \approx 2E_L$ and $\Omega
_{C}^{P-Z}\approx 4E_L$.

\begin{figure}[t]
\includegraphics[width=0.49\textwidth]{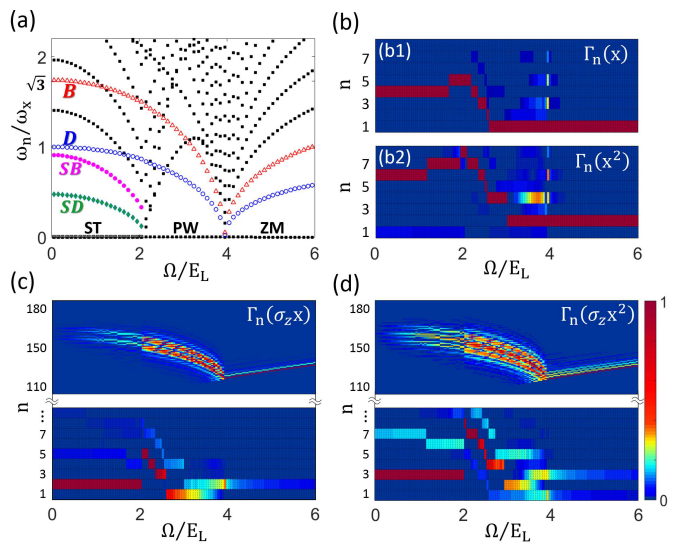}
\caption{(Color online) (a) Bogoliubov spectrum as functions of the Raman
coupling strength $\Omega$. Hollow circles and triangles denote the
frequency of the dipole (D) and the breathing (B) mode, respectively; while
solid diamonds and disks correspond to the spin dipole (SD) and the spin
breathing (SB) frequencies. Furthermore, black hollow squares label the near
zero-energy mode in the ST phase. (b1), (b2), (c) and (d) correspond to the
normalized transition strength $\Gamma_n$ of the dipole, the breathing, the
spin dipole and the spin breathing modes, respectively. (b)-(d) share the
same colormap at the right side of (d). In our calculations, we take $%
g_{\uparrow\downarrow}=0.7g$ where $g$ is calculated with the $^{87}$Rb BEC
in a condition that $\protect\omega_x=2\protect\pi \times 45\,$Hz, $%
a=101.8a_{B}$ with $a_{B}$ being the Bohr radius, and total atom number $%
N=2000$.}
\label{Fig1}
\end{figure}

\section{BOGOLIUBOV SPECTRA}

We study the collective excitation of the system using the method of
Bogoliubov theory \cite{Pitaevskii2003}. To this end, we construct the
following wave function which includes small fluctuations above the ground
state:%
\begin{equation}
\mathbf{\Psi }=e^{-i\mu t}\left[ \mathbf{\Psi }_{0}+\mathbf{u}%
_{n}\left(x\right)e^{-i\omega _{n}t}+\mathbf{v}_{n}\left(x\right)e^{i\omega
_{n}t}\right] \,,  \label{B3}
\end{equation}%
where $\mu $ is the chemical potential, $\mathbf{u}_{n}=\left( u_{n\uparrow
},u_{n\downarrow }\right) ^{T}$ and $\mathbf{v}_{n}$ $=\left( v_{n\uparrow
},v_{n\downarrow }\right) ^{T}$, satisfying the normalization condition $%
\int \mathrm{d}x\left( \left\vert \mathbf{u}_{n}\right\vert ^{2}-\left\vert
\mathbf{v}_{n}\right\vert ^{2}\right) \equiv 1$, are the Bogoliubov
quasi-particle amplitudes. By inserting $\mathbf{\Psi }$ into GP Eq.~(\ref%
{B1}) and keeping the fluctuation terms to linear order, the Bogoliubov
equations are obtained (see Appendix). We solve the Bogoliubov equations
numerically to find quasi-particle excitation frequency $\omega_n$, as well
as $\mathbf{u}_{n}$ and $\mathbf{v}_{n}$.

Figure \ref{Fig1}(a) shows a typical Bogoliubov spectrum $\omega _{n}$ as a
function of the Raman coupling strength $\Omega$, with all other parameters
fixed. First one can see that for any $\Omega$, there exists a zero mode
with $\omega_0=0$, which corresponds to the ground state itself. In the
thermodynamic limit, this zero mode corresponds to the Goldstone mode
resulting from the spontaneous breaking of the U(1) gauge symmetry. At small
$\Omega$ when the system is in the ST phase, there exists an additional
low-lying mode (marked with black hollow squares) whose frequency is very
close to zero. From the examination of the quasi-particle amplitudes, we
find that, for this mode, $-\mathbf{v}^* \approx \mathbf{u}=\mathbf{\Psi}_0^-
$ where $\mathbf{\Psi}_0^-$ is approximately given by Eq.~(\ref{B2}) with $%
C_1=-C_2$. In the thermodynamic limit, the frequency of this mode will also
vanish \cite{YLi2013,SJi2014}, and it corresponds to the second Goldstone
mode resulting from the spontaneous breaking of the translational symmetry
which is unique for the ST phase \cite{QD}. These zero modes also serve as a
self-consistency check for the accuracy of our numerical calculation.

Another rather apparent feature is that the spectrum exhibits mode softening
near the two critical Raman coupling strength where the system changes from
one phase to another. It is therefore important to classify the excitation
modes. This can be achieved by examining the corresponding quasi-particle
amplitudes $\mathbf{u}_{n}$ and $\mathbf{v}_{n}$. Specifically, we design
the following method for mode identification. First we identify the operator
$\hat{O}$ that excite a specific mode. We will focus on the following modes:
dipole, breathing, spin dipole, and spin breathing modes, with the
corresponding excitation operators $x$, $x^{2}$, $\sigma _{z}x$, and $\sigma
_{z}x^2$, respectively \cite{Stringari1996, Kimura2002, Menotti2002,
Sartori2015, Bienaime2016}. We then calculate the normalized transition
strength as follows (for details, see Appendix):
\begin{equation}
\Gamma _{n}(\hat{O})=\frac{\left\vert \left\langle n\right\vert \hat{O}%
\left\vert 0\right\rangle \right\vert }{\max \left[ \left\vert \left\langle
n\right\vert \hat{O}\left\vert 0\right\rangle \right\vert _{n=1}^{\infty }%
\right] }.  \label{B4}
\end{equation}%
Here $|0 \rangle$ denotes the ground state, and $|n \rangle$ is the $n$th
quasi-particle mode in ascending order. If a mode has $\Gamma_n(\hat{O})$
close to 1, then it reflects the main excitation features of the
perturbation $\hat{O}$, and we classify this mode accordingly. Examples of $%
\Gamma _{n}(\hat{O})$ are presented in Fig.~\ref{Fig1}(b)-(d), which help us
to identify those modes labeled in Fig.~\ref{Fig1}(a). In the following, we
present a more detailed discussion of these modes.

\section{DIPOLE AND BREATHING MODES}

A dominant dipole and a dominant breathing mode are present in all three
phases. Their normalized transition strengths are plotted in Fig.~\ref{Fig1}%
(b1) and (b2), respectively. At $\Omega=0$, i.e., in the absence of the SO
coupling, their frequencies are given by: $\omega_D=\omega_x$ \cite{Kohn}
and $\omega_B=\sqrt{3}\omega_x$ \cite{Kimura2002,Menotti2002}, in full
agreement with our numerical results. A common feature of these two modes is
that they become soft with frequency tending to zero at $\Omega=%
\Omega_{C}^{P-Z}$, i.e., at the phase boundary between the PW and the ZM
phases. However, at the other phase boundary between the ST and the PW
phase, these two modes do not exhibit any special features. The softening of
the dipole mode at $\Omega_{C}^{P-Z}$ has been studied by Li \emph{et al.}
\cite{YLi2012SumRule}. Using a sum rule approach, they have shown that near $%
\Omega=\Omega_{C}^{P-Z}$, the dipole mode frequency $\omega_D \propto 1/%
\sqrt{\chi}$ where $\chi$ is the spin polarizability which diverges at $%
\Omega=\Omega_{C}^{P-Z}$. Alternatively, as pointed out in the Refs.~\cite%
{Martone2012} and \cite{Stringari2016}, the low-lying collective modes can
be described by a hydrodynamic equation with an effective trapping frequency
$\sqrt{m/m^*}\omega_x$, and the divergence of the effective mass $m^*$
at critical point $\Omega_{C}^{P-Z}$ is also able to explain the softening
behavior in this region. In 2012, Zhang and coworkers \cite{JZhang2012}
experimentally measured the dipole oscillation frequency of an SO coupled Rb
condensate. They indeed found evidence of the mode softening near the phase
boundary. However, the frequency never reaches zero. In fact, the dipole
oscillation became quite complicated near $\Omega_C^{P-Z}$ and could not be
fitted by a single frequency. They attributed these features to the
nonlinear effects \cite{JZhang2012,YLi2012SumRule}. From our plot of the
normalized transition strength in Fig.~\ref{Fig1}(b1), we find that, near $%
\Omega_C^{P-Z}$, there exist several low-lying modes with significant dipole
transition strengths, which can be excited simultaneously in the experiment.
This can explain the complicated behavior of the dipole oscillation near $%
\Omega_C^{P-Z}$ observed in the experiment.

As we have mentioned, the breathing mode exhibits a very similar behavior as
the dipole mode and vanishes at $\Omega=\Omega_{C}^{P-Z}$. Furthermore,
there also exist several low-lying modes with significant breathing
transition strengths, as can be seen in Fig.~\ref{Fig1}(b2). Hence we expect
similar complicated behavior near the phase boundary, just as in the case of
the dipole mode.

\begin{figure}[t]
\includegraphics[width=0.40\textwidth]{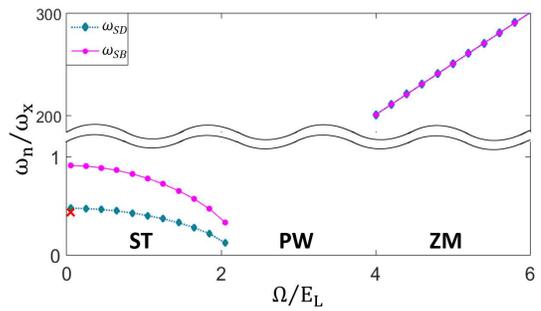}
\caption{(Color online) The spin dipole (SD) and the spin breathing (SB)
mode frequency as functions of the Raman coupling strength $\Omega$. These
two modes are not well defined in the PW phase. Other parameters are the
same as in Fig.~\protect\ref{Fig1}. }
\label{Fig2}
\end{figure}

\section{SPIN DIPOLE AND SPIN BREATHING MODES}

The spin dipole (SD) mode operator is $x \sigma_z$. In practice, this mode
can be excited by adding a spin-dependent magnetic gradient. The normalized
transition strength $\Gamma(x\sigma_z)$ is plotted in Fig.~\ref{Fig1}(c),
and the mode frequency $\omega_{SD}$ across the whole phase diagram is
plotted in Fig.~\ref{Fig2} (diamonds with dotted line), from which we
observe the following: In the ST phase, there is a dominant low-lying SD
mode. Its frequency decreases as $\Omega$ increases, but remains finite at
the ST/PW phase boundary. At $\Omega=0$, our system reduces to a binary
condensate without SO coupling. Based on the sum rule and the local density
approximation, the SD frequency for such a system is given by \cite%
{Bienaime2016}:
\begin{equation*}
\omega_{SD} = \sqrt{\frac{g-g_{\uparrow \downarrow}}{g+g_{\uparrow
\downarrow}}} \, \omega_x \,,
\end{equation*}
which leads to $\omega_{SD} \approx 0.42 \omega_x$ when plugging in the
parameters used in our calculation. This is plotted as the red cross in Fig.~%
\ref{Fig2}, and is in good agreement with our numerical result. In the PW
phase, there is no single dominant spin dipole mode. This is particularly
true away from the ST/PW phase boundary. In the ZM phase, there is a
dominant SD mode, but its frequency is much higher than the one in the ST
phase. Further examination shows that the SD frequency in this regime is
very close to the Raman coupling strength $\Omega $ (see discussion below).

\begin{figure}[t]
\includegraphics[width=0.48\textwidth]{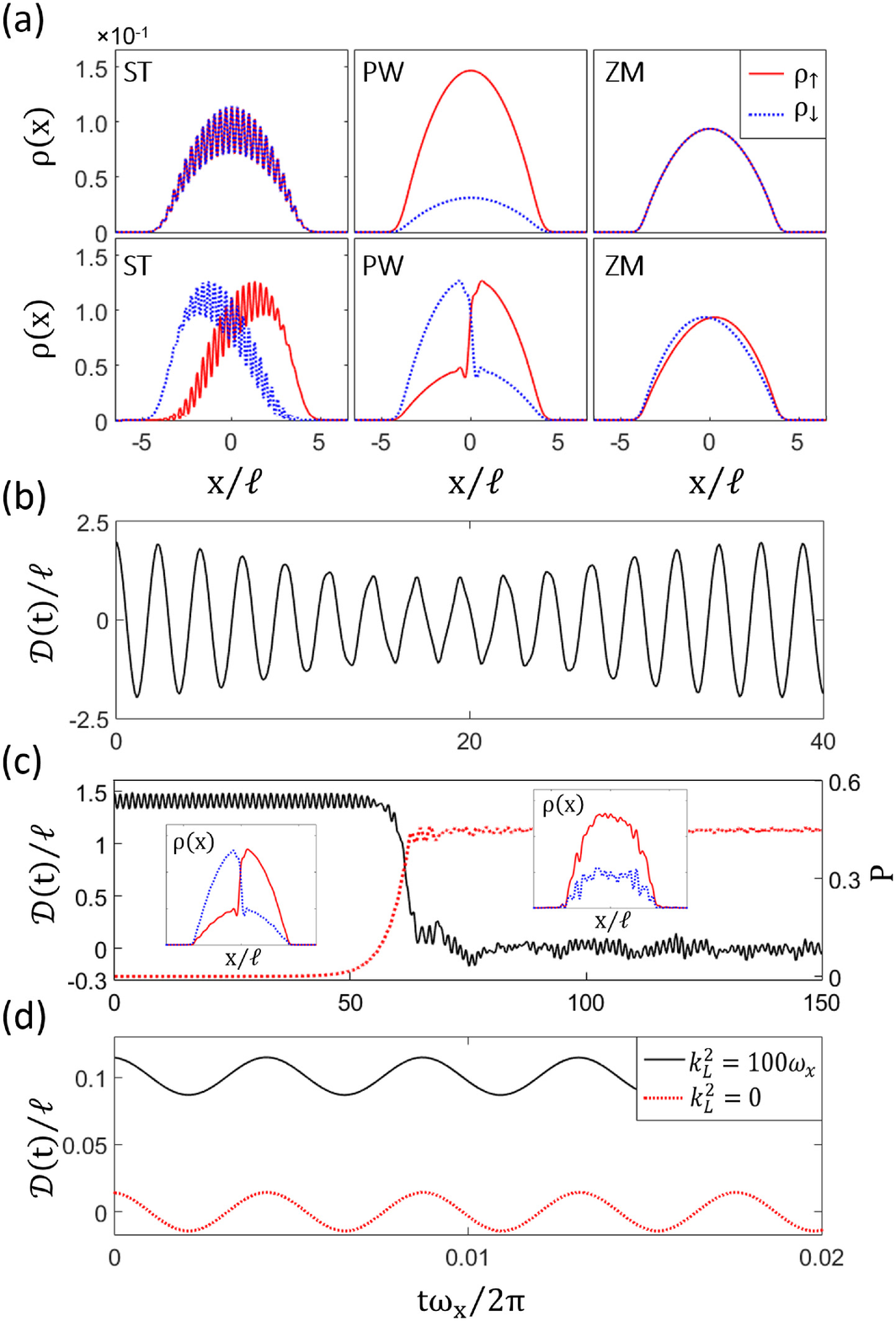}
\caption{(Color online) (a) Upper row: ground state density profiles in
three phases, where $\ell =\protect\omega_{x}^{-1/2}$ is the harmonic
oscillator length. Lower row: density profiles in the presence of a magnetic
gradient with $d=\protect\omega_{x}^{2}\ell/4$. In the calculation, we take $%
\Omega=E_L$, $3.5E_L$ and $4.5E_L$ to represent the ST, the PW and the ZM
phase, respectively. The evolution of the spin separation $\mathcal{D}(t)$
after the sudden quench of the magnetic gradient for the three phases are
illustrated in (b), (c) and (d), respectively. (b) Evolution of $\mathcal{D}%
(t)$ in the ST phase. (c) Evolution of $\mathcal{D}(t)$ (black solid line)
and polarization $P$ (red dotted line) in the PW phase. The two insets show
the typical density profiles before and after the jump which occurs around $%
t=60$ trap periods. (d) Evolution of $\mathcal{D}(t)$ in the ZM phase, with
(black solid line) and without (red dotted line) the SO coupling,
respectively. Other parameters are the same as in Fig.~\protect\ref{Fig1}.}
\label{Fig3}
\end{figure}

To gain further insights of the spin dipole mode, we carry out a direct
numerical simulation of the condensate dynamics. We add a small
spin-dependent magnetic gradient in the system, which introduces an
additional term $-d\sigma _{z}x$ in the Hamiltonian. The ground state
density profiles without and with the magnetic gradient are plotted in the
upper and lower rows in Fig.~\ref{Fig3}(a), respectively. In the absence of
the gradient, the PW phase possesses a finite magnetization and the density
profiles for the two spin components are different. By contrast, both the ST
and the ZM phases are unmagnetized with identical density profiles in the
two spin components. The characteristic density modulations in the ST phase
are, however, difficult to observe in practice, as the spatial period of
these oscillation is on the order of optical wavelength, which is far below
the resolution of a typical imaging system. This poses as a great challenge
for the direct observation of the ST phase \cite{note}. In the presence of
the gradient, the center-of-mass of the two spin components are displaced in
opposite directions. In the ST phase, the density oscillations are still
present and such oscillations in the two spin components remain in phase.
The magnetic gradient has the most dramatic effect on the PW state: the
overall magnetization is now zero but the two spin components are separated
in space with a rather sharp domain wall between them.

To study the spin dipole dynamics, we prepare the system in the ground state
in the presence of the magnetic gradient, and then suddenly quench the
magnetic gradient to zero at $t=0$ and follow the dynamics of the system by
solving the time-dependent GP equation in real time \cite{Antoine2015}. We
define the displacement between the two spin components as $\mathcal{D}%
(t)=\left\langle x_{\uparrow}(t)\right\rangle-\left\langle
x_{\downarrow}(t)\right\rangle$ with $\langle x_\sigma \rangle = \int dx
\,|\Psi_\sigma(x)|^2x$. Typical dynamics of $\mathcal{D}(t)$ for the three
phases are plotted in Fig.~\ref{Fig2}(b)-(d), which we describe below.

(1) For the ST phase depicted in Fig.~\ref{Fig3}(b), $\mathcal{D}(t)$
oscillates roughly sinusoidally around zero, with an oscillation frequency
matching very well with the one obtained from the Bogoliubov calculation $%
\omega _{\mathrm{SD}}$. Furthermore, during the time evolution, the density
modulations in the two spin components remain `phase locked'. We found that
the dynamics can be accurately reproduced using the Bogoliubov approach,
under which the time-dependent condensate wave function is given by Eq.~(\ref%
{B3}). For the SD mode, we found that $\mathbf{v}\approx 0$ and $\mathbf{u}%
\approx \eta x\mathbf{\Psi }_{0}^{-}/w$ where $\eta $ characterizes a small
excitation amplitude. From these, the density profile for each spin
component can be calculated as
\begin{eqnarray}
\rho _{\sigma }(t) &=&\left\vert \Psi _{0,\sigma }\left( x\right) +u_{\sigma
}\left( x\right) e^{-i\omega _{\mathrm{SD}}t}\right\vert ^{2}  \label{B5} \\
&\sim &\left[ \sin 2\theta \left( \cos 2k_{0}x+\frac{2\eta x}{w}\sin
2k_{0}x\sin \omega _{SD}t\right) \right.   \notag \\
&+&\left. \left( 1\pm {\frac{2\eta x}{w}\cos 2\theta \cos {\omega _{\mathrm{%
SD}}t}}\right) +O\left( \eta ^{2}\right) \right] G^{2},  \notag
\end{eqnarray}%
where $\pm $ corresponds to $\sigma =\uparrow $ and $\downarrow $,
respectively. Here, we have ignored a global normalization constant in Eq.~(%
\ref{B5}). One can see that the first term in the square bracket ensures that the
spatial density modulations are in phase for the two spin components.

(2) For the PW phase depicted in Fig.~\ref{Fig3}(c), the response of the
system is rather nonlinear. In the beginning of the evolution, $\mathcal{D}%
(t)$ carries out small-amplitude oscillations around the initial value,
indicating the presence of a stiff domain wall. At $t \approx 60$ trap
periods, the domain wall collapses and the systems jumps to a situation
close to the ground state in the absence of the gradient, accompanied by a
jump of the overall magnetization $P$ from zero to a finite value. Typical
density profiles before and after this jump are shown as the two insets in
Fig.~\ref{Fig3}(c). This nonlinear behavior is consistent with the
Bogoliubov result we obtained earlier. In particular, Fig.~\ref{Fig1}(c)
shows that there are a large number of quasi-particle modes with significant
SD transition strength in the PW phase. The dynamical behavior may be
regarded as resulting from the nonlinear mode coupling among these modes.

(3) Finally, for the ZM phase depicted in Fig.~\ref{Fig3}(d), $\mathcal{D}%
(t) $ oscillates sinusoidally with a frequency slightly above $\Omega$,
again in full agreement with the Bogoliubov result. Furthermore, $\mathcal{D}%
(t)$ never changes sign for the parameters we used in our simulation (if we
use a large enough $\Omega$, $\mathcal{D}(t)$ may change sign, but the
oscillation would remain asymmetric about zero). In comparison, we also
simulated the spin dipole dynamics of a coherently coupled two-component BEC
without SO coupling \cite{Sartori2015}, by taking $k_L=0$ in Hamiltonian~(%
\ref{h0}), shown as the red dotted line in Fig.~\ref{Fig3}(d). The main
difference with and without the SO coupling is that, in the latter case, $%
\mathcal{D}(t)$ oscillates symmetrically about zero. This difference can be
understood in a simple way. The Raman coupling term in Hamiltonian~(\ref{h0}%
) may be regarded as an effective uniform transverse magnetic field along
the $x$-axis. In the absence of the magnetic gradient, the atoms are
therefore spin polarized along the $x$-axis. The weak magnetic gradient tips
the atomic spin slightly away from the $x$-axis. After the quench of the
magnetic gradient and in the absence of the SO coupling, the atomic spin
precesses around the effective transverse magnetic field, with the Larmor
frequency given by $\Omega$. In one period, the spin rotates about the $x$%
-axis in a full circle, rendering $\mathcal{D}(t)$ to oscillate about zero
symmetrically. By contrast, in the presence of the SO coupling, the SO
coupling term may be regarded as an effective momentum-dependent
longitudinal magnetic field along the $z$-axis, which tends to maintain the
value of $\sigma_z$. Hence the spin flip becomes incomplete and $\mathcal{D}%
(t)$ tends to maintain its original sign.

Finally, let us briefly discuss the spin breathing (SB) mode with the
corresponding perturbation operator $\sigma_z x^2$. This mode can be excited
by adding a spin-dependent trapping potential such that the two spin
components experience different trapping frequencies. The normalized SB
transition strength $\Gamma_{n}(\sigma x^2)$ is plotted in Fig.~\ref{Fig1}%
(d), and the mode frequency $\omega_{SB}$ across the whole phase diagram is
plotted in Fig.~\ref{Fig2} as circles with solid line. The behavior of $%
\omega_{SB}$ as a function of $\Omega$ is very similar to that of $%
\omega_{SD}$ we discussed above. As a result, we do not present a detailed
discussion about this mode here.

\section{CONCLUSION}

To summarize, we have presented detailed study of the collective excitation
properties of a quasi-1D SO coupled BEC. This system possesses three phases
which can be accessed by tuning the Raman coupling strength. We developed a
method to efficiently classify the numerically obtained Bogoliubov
excitation modes. We show that the dipole and the breathing modes become
soft at the boundary between the PW and the ZM phases, but are smooth across
the boundary between the ST and the PW phases. By contrast, the spin dipole
and the spin breathing modes have distinct features in all three phases. We
hope that our work may stimulate more experimental study of the collective
excitation properties of SO coupled BEC.

\bigskip

\bigskip

\bigskip

\begin{acknowledgments}
L. Chen would like thank T.-T. Li for helpful discussion. YZ is supported by
NSF of China under Grant Nos. 11234008 and 11474189, the National Basic
Research Program of China (973 Program) under Grant No. 2011CB921601,
Program for Changjiang Scholars and Innovative Research Team in University
(PCSIRT)(No. IRT13076). ZQY is supported by NSFC under Grant No. 11674202.
HP acknowledges support from US NSF and the Welch Foundation (Grant No.
C-1669).
\end{acknowledgments}

\appendix

\section{BOGOLIUBOV EQUATIONS AND NORMALIZED TRANSITION STRENGTH}

In this Appendix, we provide more details on how the Bogoliubov spectrum and
the normalized transition strength are calculated. The Bogoliubov equations
are in the form of
\begin{eqnarray}  \label{A1}
\omega \mathbf{u} &=& \left(H_{0}+A\right)\mathbf{u}+B\mathbf{v}\,, \\
-\omega \mathbf{v} & =&\left(H_{0}^{*}+A^*\right)\mathbf{v}+B^*\mathbf{u}\,,
\end{eqnarray}
where
\begin{equation}
A=\left(
\begin{array}{cc}
2g\rho _{0,\uparrow }+g_{\uparrow \downarrow }\rho _{0,\downarrow }-\mu &
g_{\uparrow \downarrow }\psi _{0,\uparrow }\psi _{0,\downarrow }^{\ast } \\
g_{\uparrow \downarrow }\psi _{0,\uparrow }^{\ast }\psi _{0,\downarrow } &
2g\rho _{0,\downarrow }+g_{\uparrow \downarrow }\rho _{0,\uparrow }-\mu%
\end{array}%
\right) \,,  \notag
\end{equation}%
and%
\begin{equation*}
B=\left(
\begin{array}{cc}
g\psi _{0,\uparrow }^{2} & g_{\uparrow \downarrow }\psi _{0,\uparrow }\psi
_{0,\downarrow } \\
g_{\uparrow \downarrow }\psi _{0,\uparrow }\psi _{0,\downarrow } & g\psi
_{0,\downarrow }^{2}%
\end{array}%
\right) \,,
\end{equation*}%
are related to the two-body interaction, $\rho _{0,\sigma }=\left\vert \psi
_{0,\sigma }\right\vert ^{2}$ are the ground-state density, and $\mu $
denotes the chemical potential. In our numerical calculation, we first
obtain the ground-state wave function $\mathbf{\Psi }_{0}$ by propagating
the Gross-Pitaevskii equations, Eq.~(\ref{B1}) in the main text, in
imaginary time. And then, the Bogoliubov spectrum $\omega _{n}$ as well as
the amplitudes $\mathbf{u}_{n}$ and $\mathbf{v}_{n}$ can be worked out by
directly diagonalize the Bogoliubov equations using the Arnoldi method \cite%
{Arnoldi}.

In the framework of linear response theory, the total field operator $\mathbf{%
\hat{\Psi}}$ can be linearized into the form of
\begin{equation}
\mathbf{\hat{\Psi}}\left( x\right) =\mathbf{\Psi}_{0}+\sum_{n=1}
\begin{array}{c}
\mathbf{u}_{n}\hat{b}_{n}+\mathbf{v}_{n}^{\ast }\hat{b}_{n}^{\dagger }%
\end{array}%
,  \label{BB1}
\end{equation}%
where $\hat{b}_{n}$ is the annihilation operator for the $n{\rm th}$ quasi-particle state $|n \rangle$. The
transition strength $\Gamma _{n}(\hat{O})$ can be obtained by calculating the matrix element
\[\langle n|\hat{O} |0 \rangle \equiv \int dx\,\left\langle n\right\vert  (\mathbf{\hat{\Psi}}^{\dag }\hat{O}\mathbf{%
\hat{\Psi}}) \left\vert 0\right\rangle \,.\] Specifically, for the
spin-independent operators $\hat{O}=x^{n^{\prime }}$, we have
\begin{align}
\left\langle n\right\vert x^{n^{\prime }}\left\vert 0\right\rangle =\int
&dx(\psi _{0,\uparrow }x^{n^{\prime }}u_{n,\uparrow }^{\ast }+\psi
_{0,\uparrow }^{\ast }x^{n^{\prime }}v_{n,\uparrow }^{\ast }  \notag \\
&+\psi _{0,\downarrow }x^{n^{\prime }}u_{n,\downarrow }^{\ast }+\psi
_{0,\downarrow }^{\ast }x^{n^{\prime }}v_{n,\downarrow }^{\ast })\,,
\label{BB2}
\end{align}%
while for the spin-dependent operators $\hat{O}=\sigma _{z}x^{n^{\prime }}$,
we have%
\begin{align}
\left\langle n\right\vert \sigma _{z}x^{n^{\prime }}\left\vert
0\right\rangle =\int &dx(\psi _{0,\uparrow }x^{n^{\prime }}u_{n,\uparrow
}^{\ast }+\psi _{0,\uparrow }^{\ast }x^{n^{\prime }}v_{n,\uparrow }^{\ast }
\notag \\
&-\psi _{0,\downarrow }x^{n^{\prime }}u_{n,\downarrow }^{\ast }-\psi
_{0,\downarrow }^{\ast }x^{n^{\prime }}v_{n,\downarrow }^{\ast })\,.
\label{BB3}
\end{align}%
Again, we keep small amplitudes $\mathbf{u}_{n}\left( x\right) $ and $%
\mathbf{v}_{n}\left( x\right) $ into the linear term in Eqs.~(\ref{BB2}) and
(\ref{BB3}). Here $n^{\prime }=1$
and $2$ correspond to the (spin) dipole and the (spin) breathing operators,
respectively.

\end{document}